\documentclass[preprint,12pt]{elsarticle}
\usepackage{amssymb}

\journal{Physics Letters B}

\newcommand\be{\begin{equation}}
\newcommand\ee{\end{equation}}
\newcommand\bea{\begin{eqnarray}}
\newcommand\eea{\end{eqnarray}}
\newcommand\apj{Astrophys. J.}
\newcommand\apjl{Astrophys. J. Letts.}

\newcommand\pasa{Publ. Astron. Soc. Aust.}

\begin{document}

\begin{frontmatter}

\title{Collision of ultra-relativistic proton with strong magnetic field:
production of ultra-high energy photons and neutrinos}
\author{Ye-Fei Yuan and Xin-Yue Shi}
%\email[]{yfyuan@ustc.edu.cn}
\address{CAS Key Laboratory for Research in Galaxies and Cosmology,
Department of Astronomy, University of Science and Technology of China,
Hefei 230026, China\\
School of Astronomy and Space Science, University of Science and Technology
of China, Hefei 230026, China}

\begin{abstract}
Proton-proton interaction and photo-hadronic interaction in cosmic accelerators
are the two main channels for the production of cosmic ultra-high energy
photons and neutrinos (TeV-PeV).
In this {\it Letter}, we use FWW approach to obtain the production of cosmic
ultra-high energy photons and neutrinos from the collision between
UHE proton with magnetic field which could be considered as the virtual
photon in the rest frame of UHE proton. We name this as $pB$ process.
The threshold for the occurrence of the $pB$ process is that the combination
of the Lorentz factor of proton and the strength of the magnetic field is
about $\gamma_p B \simeq 5\times 10^{18}$Gauss. Beyond this threshold,
the rate of energy loss of proton due to the $pB$ process is about
three orders higher than that due to the synchrotron radiation of proton
in the same magnetic field.
The $pB$ process might potentially happen in the atmosphere of white dwarfs,
neutron stars or even that of stellar massive black holes.
\end{abstract}


\end{frontmatter}

\section{Introduction}

Six years after its completion, {\it IceCube} has detected more than
80 neutrinos with energies between 100 TeV and 10 PeV
\citep{Aartsen2014}\footnote{https://icecube.wisc.edu}. For the production of ultra-high energy
(UHE) neutrinos whose energy is above TeV, generally a
cosmic accelerators for UHE protons and a target which could be
either protons or radiation field are needed. In this {\it Letter},
we suggest that magnetic field could be another target.
As seen by an UHE proton,
the static magnetic field is equivalent to electromagnetic field,
the collision between the UHE proton and the equivalent
electromagnetic field (virtual photons) could produce UHE pions
($\pi^{0,\pm}$), decay of pions produce UHE photons and neutrinos,
which is very similar to the photo-hadronic interaction.
We name the collision of the UHE proton with magnetic field as
the ``$pB$" process.

The original idea of virtual quanta
is from Fermi (1924), Weizsacker (1934), and Williams (1935),
who proposed that the static electric field of
nuclei as seen by a relativistic electron is equivalent to the
electromagnetic field, the scattering of the electromagnetic field
by the relativistic electron produces the so called bremsstrahlung
radiation \citep{Fermi1924, Weizsaecker1934, Williams1935, Jackson1998}.
The Fermi-Weizsacker-Williams (FWW)
method was also successfully applied to unify the synchrotron radiation
and the inverse Compton scattering\citep{Lieu1993}.
Applying the FWW method, Zhang \& Yuan (1998) obtained the quantum-limited
synchro-curvation radiation formulae, which is very difficult to obtain in
quantum electrodynamics \citep{Zhang1998}. Only recently, there appears
an attempt to derive the quantum theory of synchro-curvation radiation from
first principles \citep{Voisin2017a,Voisin2017b}.

As protons strongly couple to meson field, proton synchrotron emission in a strong magnetic field
could also produce pions and other mesons, similar to the production of photons.
$\pi^{0}$ and other heavy meson productions of proton synchrotron emission have been studied in semi-classical approximations\cite{Tokuhisa1999, Berezinsky1995}.
This process is also calculated in a fully relativistic and quantum-mechanical way\cite{Maruyama2015} including the effect of proton anomalous magnetic moment in a strong magnetic field of $10^{18}$ G.
Using scaling relations\cite{Maruyama2016}, the decay width can be extrapolated to the proton energy $\sim$ TeV in magnetic fields of $10^{15}$ G, where the Landau numbers of the initial and final protons are $n_{i,f} \sim 10^{12}$-$10^{13}$.

However, in the above quantum-mechanical analysis of the
pion($\pi^0$) production from ultra-relativistic proton in strong
magnetic field, only one channel for the production of $\pi$ meson
are considered:
$p+B \rightarrow p+\pi^0$ (direct-pion production).
Actually, there are many channels for the
photo-hadronic interaction, such as, $\Delta$-resonance, $N$-resonance,
direct-pion production and multi-pion production,
and the dominate process is $\Delta$-resonance
(\cite{1999PASA...16..160M, Muecke2000}). And the quantum-mechanical
calculations is only valid for the case of the strong uniform straight
magnetic field. If not,
it is very difficult to solve the Dirac equation of proton in
a curve magnetic field.
Therefore, in this {\rm letter} we discuss the $\pi^{0,\pm}$ production
from the point view of FWW approach.
There are several advantages in our method. First,
the contributions from all channels for the production
of  $\pi^{0,\pm}$ are considered by using the total $p\gamma$ photon-meson
cross section (for example, see Figure 1
in \cite{1999PASA...16..160M,Huemmer2010}). Second,
the production rate for $\pi^{\pm}$ is also provided in our method,
the decay of $\pi^{\pm}$ is one of the main production mechanisms
of ultra high energy neutrinos in astrophysical sources.
Third, according to our argument, our results are
also valid in curved magnetic fields, even chaotic magnetic fields,
provided that the Larmor radius of proton is less than
the curvature radii of the local magnetic field\cite{Lieu1995}.
Finally, our method is physical transparent, which is very helpful for
understanding the results obtained in more theoretical calculations.

\section{Semi-quantitative analysis}
As a first step, a semi-quantitative analysis of the $pB$ process is presented
to show the basic idea of  ``$pB$" process via the FWW approach,
the detailed analysis can be found in the following discussion.

Suppose there is a uniform straight magnetic field $B$ along z-axis direction
in the laboratory frame $\Sigma$.
Now an UHE proton with Lorentz factor $\gamma_p$
is colliding with the magnetic field.
Ignore the proton velocity in the z-axis,
the electromagnetic field in the proton instantaneous rest frame
$\Sigma'$ can be estimated via the Lorentz transformation to be
\be
E'=\gamma_{ p} B \left( \frac{v_y}{c},\,\,\,-\frac{v_x}{c},\,\,\,0 \right);
B'=\gamma_{ p} B(0,\,\,\,0,\,\,\,1),
\ee
where $\vec{v}$ is the velocity of proton in the laboratory frame
$\Sigma$. For ultra-relativistic particle,  $\gamma_p \gg 1$, and
$v \rightarrow c$, so the effect of passing segment
of the field resembles an electromagnetic wave propagating
in the $-\vec{v}$ direction, as seen by the observer in the
instantaneous rest frame of proton.
The apparent Poynting energy flux $S'$ in the rest frame is
$\sim c \gamma_{p}^2 B^2/(4 \pi)$.

First, collision of the UHE proton with the magnetic field lead to the
energy loss of the proton via the synchrotron radiation.
Using the FWW approach, we can obtain the total energy loss rate
of the proton synchrotron radiation,
which is given by the product of the Poynting energy flux $S'$
and the proton Thomson cross section $\sigma_{\rm T}^p = 8 \pi e^4 /(3m_{p}^2 c^4)$, where $m_p$ is the mass of proton, viz.,
\be
P_{\rm syn} \simeq \sigma_{\rm T}^p S' \simeq \frac{2\gamma^2 e^4 B^2}{3m_p^2c^3} \simeq
\sigma_{\rm T}^p c \gamma_{ p}^2 U_{\rm B},
\label{eq:simsyn}
\ee
here $U_{\rm B} = B^2/(8\pi)$ is the energy density of magnetic fields in the laboratory frame.
This is the standard result for classical synchrotron radiation \cite{Lieu1993}.

If we replace the Thomson cross section $\sigma_{\rm T}^p$
with the cross section of photo-meson interaction,
we can estimate the energy loss rate of the $pB$ process.
It is well known that the cross section of the photon-meson interaction is
energy-dependent. For simplicity, we first consider the resonant interaction,
which dominates the process of photo-meson interaction. The cross section
of the resonant interaction is
$\sigma_{\rm res} \simeq 200 \mu {\rm barn}$
for the photons with the energy between 0.2 GeV and 0.5 GeV in the rest frame
of proton.
The total rate of the energy loss of the UHE proton in the resonant region is
\be
P_{pB} \simeq \sigma_{\rm res} S' = \sigma_{\rm res} c \gamma_{ p}^2 U_{\rm B}.
\label{eq:simres}
\ee

In the FWW approach, the typical frequency of the equivalent electromagnetic field
is the cyclotron frequency $\omega'_{\rm c}$ of proton in its rest frame,
\be
\omega'_{\rm c} = \frac{\gamma_{p} e B}{m_{p} c}.
\ee
Therefore, for the occurrence of the $pB$ process, the energy of the equivalent
photon should be in the range of 0.2 GeV and 0.5 GeV, i.e.
\be
0.2 {\rm GeV} < \hbar \omega'_c
= \frac{\gamma_{p} \hbar e B}{m_{p} c} <0.5 {\rm GeV},
\ee
which requires the product of proton Lorentz factor $\gamma_p$ and strength of magnetic field in the range of
\be
0.2B_{\rm c}^{ p} < \gamma_{ p} B < 0.5B_{\rm c}^{ p}. \label{eq:B}
\ee
Here $B_{\rm c}^{ p}=m_{ p}^2 c^3/(\hbar e) =1.5\times 10^{20}$G
is the critical magnetic field of proton.

The Thomson cross section of proton is $\sigma_{\rm T}^p \approx 0.20 \mu {\rm barn}$,
compare with Eq.(\ref{eq:simsyn}) and Eq.(\ref{eq:simres}),
it is evident that the $pB$ process dominates
the energy loss of proton, when the $pB$ process switches on.

%\be
%3.2 \times 10^{19} {\rm G} \simeq 0.2B_{\rm c}^{ p} < \gamma_{ p} B < 7.9 \times 10^{19} {\rm G} \simeq 0.5B_{\rm c}^{p}
%\ee

\section{Proton magnetic field process}

%%\subsection{Classical theory of equivalent photon}
In the following, more detailed derivation of the
rate of the energy loss of UHE proton, and relevant emissivity of
UHE neutrinos are given.
As well known, a relativistic proton moves in a circular orbit
in the laboratory frame $\Sigma$,
\be
x=R\sin \omega_0 t ,\,\,\, y=R \cos \omega_0 t ,
\ee
where $R$ is the radius, $\omega_0 = eB/\gamma_p m_p c$ is the cyclotron frequency of
proton in the laboratory frame $\Sigma$.
The proton velocity in the instantaneous rest frame $\Sigma '$ is
\bea
v'_x &=& \frac{dx'}{dt'}=\frac{dt}{dt'}\frac{dx'}{dt}=-\frac{v(1-\cos \omega _0 t)}{1-\frac{v^2}{c^2} \cos \omega _0 t} \label{eq:motionx} \\
v'_y &=& \frac{dy'}{dt'}=\frac{dt}{dt'}\frac{dy'}{dt}=-\frac{v\sin \omega _0 t}{\gamma(1-\frac{v^2}{c^2} \cos \omega _0 t)} \label{eq:motiony}
\eea

\begin{figure}[htpb]
\centering
\includegraphics[width=0.3\textwidth]{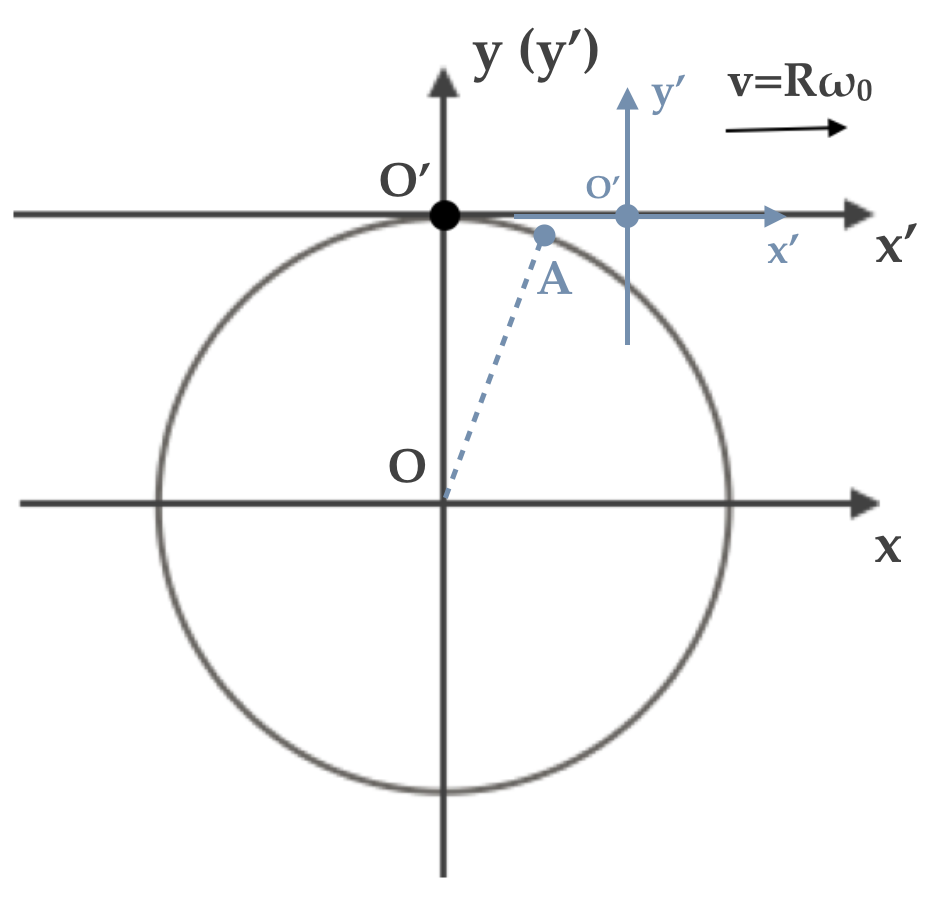}
\caption{Schematic shows a segment of the trajectory of proton
in the laboratory frame. When $t=0$, proton is at point $O'$,
the origin of $\Sigma'$ frame at $t=0$,
and when $t=\Delta t$, proton moves
to point A along a circular orbit. During the period of the movement
of proton, the inertial rest frame of proton $\Sigma'$
moves to a new place along $x'$-axis with a constant velocity
$v=R\omega_0$.}
\end{figure}

As it is shown in Fig.1, when $t=\Delta t$, the proton has a deviation
from the origin point $O'$ in the its own rest frame $\Sigma'$.
In the regime of the classical radiation, which means that
the motion of particle is classical, and the Landau quantization
is unimportant, the acceleration of proton in its own rest frame
$\Sigma'$ is due to the force of equivalent electric field $E'(t')$.
Moreover, using the Fourier analysis of the equivalent electric
field $E'(t')$, we can obtain the spectral Poynting flux $S'(\omega ')$
(energy per unit area per unit frequency) of the equivalent
incident radiation, which is given by \citep{Lieu1993},
\bea
S'(\omega')
&=& c \left| \frac{1}{2\pi}\int_{-\infty}^{\infty}{E'e^{i\omega't'}dt'}
	\right| ^2 \nonumber  \\
&=& c \left( \frac{\omega' m_p }{2\pi e} \right)^2 \left|
	\int_{-\infty}^{\infty}{v'_{\perp}(t')e^{i\omega't'}dt'} \right| ^2 ,
\eea
This flux can be decomposed into two polarization modes (1,0,0) and (0,1,0),
\be
S'(\omega')=c\left(\frac{\omega' m_p }{2\pi e}\right)^2 \left(|I'_x|^2+|I'_y|^2 \right), \label{polar}
\ee
where
\bea
I'_x &=&\int_{-\infty}^{\infty}{v'_{x}(t')
	e^{i\omega't'}dt'}, \label{eq:Ix}\\
I'_y &=&\int_{-\infty}^{\infty}{v'_{y}(t')
	e^{i\omega't'}dt'}. \label{eq:Iy}
\eea

Substitute Eq.(\ref{eq:motionx})-(\ref{eq:motiony})
into Eqs.(\ref{eq:Ix})-(\ref{eq:Iy}), and make
the change of variable $t' \rightarrow t$,
and ignore terms of order $\lesssim 1/\gamma_{ p}^2$,
the integral in Eqs.(\ref{eq:Ix})-(\ref{eq:Iy}) can be obtained in terms of
the modified Bessel functions (see also ref.\citep{Jackson1998, Lieu1989}),
\bea
I'_x
&\approx& \frac{\sqrt{2} v }{\omega_0 \gamma_{ p}^2}
	\int_{-\infty}^{\infty} e^{i \frac{3}{2} \xi \left(a+\frac{1}{3} a^3
	\right)} da \nonumber \\
&=& \frac{2\sqrt{6}}{3} \left( \frac{c}{\omega_0 \gamma_{ p}^2} \right)
	K_{1/3}(\xi) \\
I'_y
&\approx& \frac{2v }{\omega_0 \gamma_{ p}^2} \int_{-\infty}^{\infty} a
	e^{i \frac{3}{2} \xi \left(a+\frac{1}{3} a^3 \right)} da \nonumber \\
&=& \frac{4\sqrt{3}}{3} \left( \frac{c}{\omega_0 \gamma_{ p}^2} \right)
K_{2/3}(\xi) ,
\eea
where,
\be
a=\frac{\sqrt{2}}{2} \gamma_{ p} \omega_0 t ,\,\,\ \xi=\frac{2 \sqrt{2} \omega'}{3 \gamma_{ p}^2 \omega_0} .
\ee
From the properties of modified Bessel functions, it is evident that
the flux peaks at $\xi \simeq 1$, that is,
\be
\omega'_{\rm peak} \simeq \gamma_{ p}^2 \omega_0
	=  \frac{\gamma_{ p} e B}{m_{ p} c} = \omega_c'
\ee
Eventually, the Poynting flux in the instantaneous rest frame of proton reads,
\be
S'(\omega')
= \frac{2 m_{ p}^2 c^3}{3 \pi^2 e^2}
	\left(\frac{\omega'}{\omega_0 \gamma_{ p}^2} \right)^2
	\left[K_{1/3}^2(\xi)+ 2 K_{2/3}^2(\xi) \right] .
\ee
In order to get the power spectrum of the virtual photon
$d^2 P'/d \epsilon^{'}_{\rm r} dA^{'}$ (energy per unit time, per unit area
and per unit photon energy), the Poynting flux should be divided by
the cyclotron period in the instantaneous frame $2 \pi/ (\gamma_{ p}^2 \omega_0)$,
%\be
%F'(\epsilon_{\rm r}, \gamma_{ p}, B) \approx 3 \times 10^{27} \ {\rm cm^{-2}}\ \gamma_{ p} \omega_0 \left(\frac{\epsilon_{\rm r}}{\hbar \omega_0 \gamma_{ p}^2}\right)^2 \left[K_{1/3}^2 \left(\frac{2 \sqrt{2} \epsilon_{\rm r}}{3 \gamma_{ p}^2 \hbar \omega_0} \right)+ 2 K_{2/3}^2 \left(\frac{2 \sqrt{2} \epsilon_{\rm r}}{3 \gamma_{ p}^2 \hbar \omega_0} \right)\right] ,
%\ee
\bea
\frac{d^2 P'}{d \epsilon'_{\rm r} dA'}
&=&
\frac{m_{ p} c^2 \gamma_p B}{3 \pi^3 e\hbar}
\left(\frac{\omega'}{\omega_0 \gamma_{ p}^2} \right)^2
\left[K_{1/3}^2(\xi)+ 2 K_{2/3}^2(\xi) \right]
	\nonumber \\
&\simeq & 4.7 \times 10^{51}{\rm cm}^{-2}{\rm s}^{-1}
	\left( \frac{ \gamma_{ p} B}{B_{\rm c}^{ p}} \right)^{-1}
\left(\frac{\epsilon^{'}_{\rm r}}{m_{ p} c^2} \right)^2
	\nonumber \\
&&\left[K_{1/3}^2 (\xi)
	+ 2 K_{2/3}^2(\xi) \right]
\eea
where $\epsilon^{'}_{\rm r}$ is the energy of virtual photon in the
particle instantaneous frame, and $\xi$ could be re-expressed as
\be
\xi = \frac{2\sqrt{2}}{3}
        \left(\frac{\gamma_{ p} B}{B_{\rm c}^{p}} \right)^{-1}
        \left(\frac{\epsilon^{'}_{\rm r}}{m_{ p}c^2}\right)
\ee
When $\omega' \sim \omega'_{\rm peak}$, the power spectrum of virtual photons
reaches its peak flux of
\be
\frac{d^2 P'}{d \epsilon^{'}_{\rm r} dA^{'}}
\sim \frac{(\gamma_p B)^2 c}{4 \pi^2 \hbar \omega'_{\rm peak}} ,
\ee
which is in a good agreement with the result from
the semi-quantitative analysis.

%%\subsection{Proton energy loss rate}

Based on the spectrum of incident equivalent photons,
the rate of the energy loss of proton can be estimated
as follow,
\be
P_{{\rm p},{\rm loss}}(E_{p})
=\sum_{\rm IT} \left[ K^{\rm IT} E_{p} \Gamma^{\rm IT}(E_{p}) \right],
\ee
where IT stands for the types of interaction, $K^{\rm IT}$ is
the averaged ratio of the energy loss of proton after the
corresponding interaction, and $\Gamma^{\rm IT}$ is the rate of
interaction.
The photo-meson interaction can be classified into three types:
the resonant, direct, and multi-pion production.
We use the simplified model B (Sim-B) in H$\rm{\ddot{u}}$mmer (2010)
\citep{Huemmer2010}, and concentrate on the resonances first.
The total cross section of the resonances is dominated by
the $\Delta$(1232)-resonance (LR) at low energy, and the
higher resonances (HR) at high energy.
The parameters for both resonances are shown in Table.\ref{tab:res},
and the parameters for the other types of interaction can be found
in H$\rm{\ddot{u}}$mmer (2010).

In this study, the interaction rate can be written as
\be
\Gamma^{\rm IT}(E_{p}, B)=
\gamma_{p}^{-1} \int_{\epsilon^{\rm IT}_{\rm min}}^{\epsilon^{\rm IT}_{\rm max}}  \epsilon_{\rm r}^{-1} \frac{d^2 P'}{d \epsilon^{'}_{\rm r} dA^{'}}(\epsilon_{\rm r}, \gamma_{p}, B) \sigma^{\rm IT} (\epsilon_{\rm r}) d\epsilon_{\rm r},
\ee
where $\epsilon^{\rm IT}_{\rm min}$ and $\epsilon^{\rm IT}_{\rm max}$ show
the energy range of photons ($\epsilon_{\rm r}$) for the occurrence of the
resonance interactions, $\sigma^{\rm IT} (\epsilon_{\rm r})$ is the
total cross section of the corresponding resonant interaction
in the proton rest frame.

\begin{table}[]
\centering
\begin{tabular}{ccccc}
\hline
IT & $\ \epsilon_{\rm min}$ [GeV] & $\ \epsilon_{\rm max}$ [GeV] & $\ \sigma$ [$\mu$ barn] & $K$   \\
\hline
LR & 0.2                          & 0.5                          & 200                     & 0.22 \\
HR & 0.5                          & 1.2                          & 90                      & 0.39 \\
\hline
\end{tabular}
\caption{Parameters for the $\Delta$-resonance (LR) and the
higher resonance (HR). The range of energy of photons for the
occurrence of resonance is determined by $\epsilon_{\rm min}$
and $\epsilon_{\rm max}$, $\sigma$ is the total cross section,
and $K$ is the ratio of energy loss of proton
\citep{Huemmer2010}.}
\label{tab:res}
\end{table}

\begin{figure}[htpb]
\centering
\includegraphics[width=0.4\textwidth]{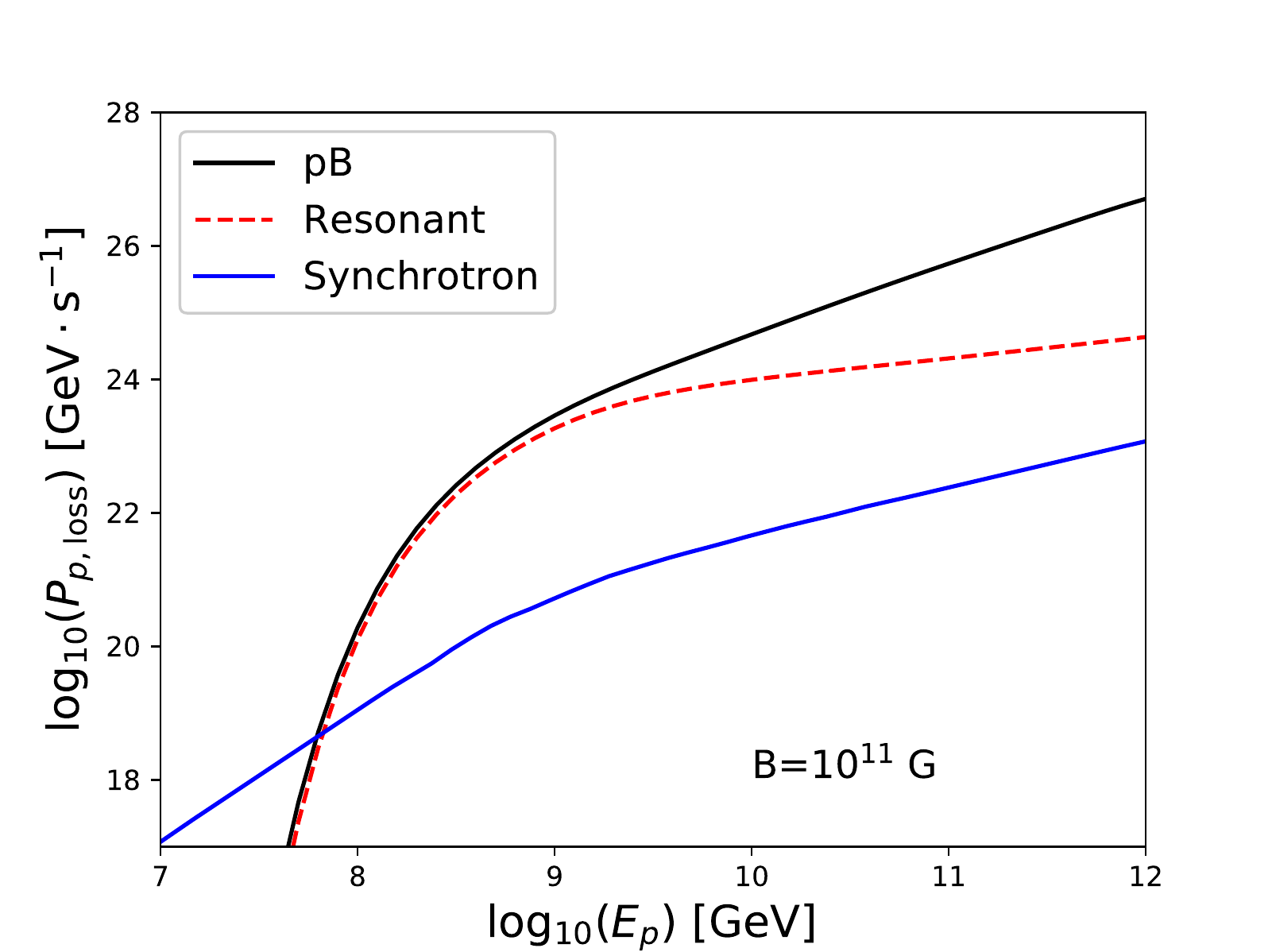}
\caption{The rate of energy loss of ultra-relativistic proton as a
function of its energy. The total rate due to the $pB$ process are shown
with the black solid line, and that due to the synchrotron radiation
with the blue solid line, while the red dashed line shows the contribution
from the resonant interaction. The strength of the magnetic field is
taken to be $B=10^{11}$G.
}
\label{fig:diffcross}
\end{figure}

\begin{figure}[htpb]
\centering
\includegraphics[width=0.4\textwidth]{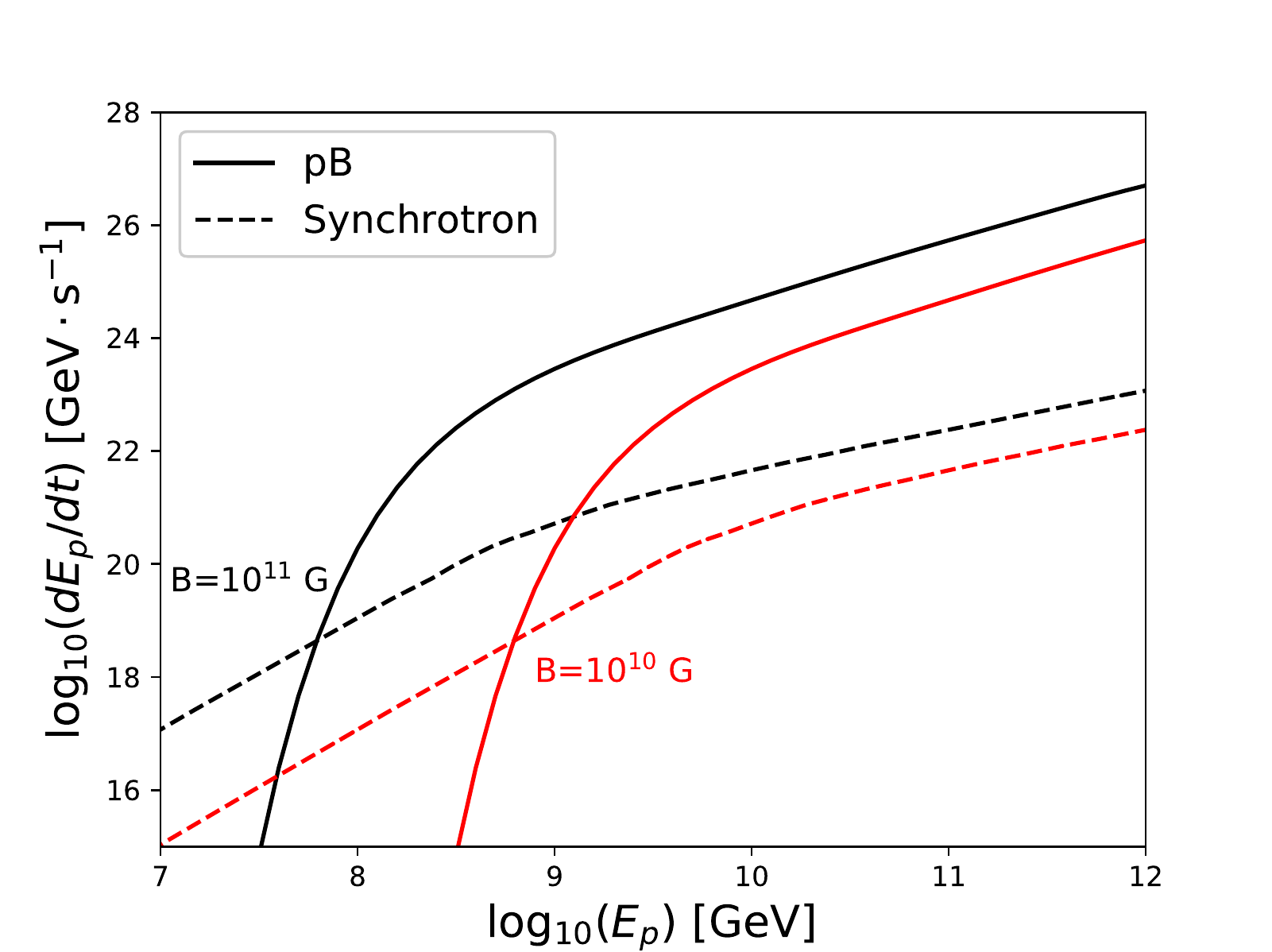}
\caption{The rate of energy loss of ultra-relativistic proton as a
function of its energy in different magnetic field:
$B=10^{10}$G (red lines) and $B=10^{11}$G (black lines).
Solid lines show the rates due to the  $pB$ process, while
the dashed lines show the rates due to the synchrotron radiation.
}
\label{fig:lossp}
\end{figure}

Figure \ref{fig:diffcross} shows rates of energy loss of proton
in the magnetic field with strength of $B=10^{11}$ G.
It is clearly shown in the figure that as the increase of the
energy of proton, the rate of energy loss due to the  $pB$ process
increases dramatically, and dominates the loss of energy. This happens
at $\gamma_{\rm p} \sim 10^8$, which is consistent with Eq.(\ref{eq:B}).
To be more realistic,
the contributions of the rate of energy loss due to the direct and
multi-pion productions are also shown in Fig.~\ref{fig:diffcross}.
In the range of low energy, the resonances dominate the loss of energy,
and the multi-pion production dominates the energy loss
in the range of high energy.

%The comparison of the proton energy loss rate via the $pB$ process
%and synchrotron radiation in different magnetic field strengths

The rates of energy loss of proton in the magnetic field with
different strength are shown in Fig.\ref{fig:lossp}.
When $\gamma_{p} B \gtrsim 5 \times 10^{18}$ G, the rate due to
the $pB$ process exceeds that due to the synchrotron radiation,
and the $pB$ process dominates the energy loss of proton.
In the resonances dominated region, the rate due to
the $pB$ process is $\sim$
three orders greater than that due to the synchrotron radiation.
Considering the broadened spectrum of incident equivalent
photons, these results are consistent with
our semi-quantitative analysis.

The $pB$ process produces the secondary gamma-ray photons and
neutrinos, which resembles the conventional $p\gamma$ process.
Non-thermal TeV-PeV photons can result from the
decay of $\pi^0$ ($\pi^0 \rightarrow \gamma + \gamma$).
The Universe is not transparent to these photons from the cosmic
distance, tens of TeV
photons can annihilate with the cosmic background radiation
into electron/positron pairs.
Therefore, generally only softer and fainter photon
spectrum would be observed.

However, the high-energy neutrinos can reach the Earth
without annihilation.
The total power of neutrino emission can be estimated
as $P_{\nu} = \frac{3}{4} P_{p,{\rm loss}}$, where the
factor $\frac{3}{4}$ is the averaged energy fraction
transferred from pion to neutrinos in the decay chains
of $\pi^{+} \rightarrow \nu_{\mu} \mu^{+}
\rightarrow \nu_{\mu} e^{+} \nu_{e} \bar{\nu}_{\mu}$
and $\pi^{-} \rightarrow \bar{\nu}_{\mu} \mu^{-}
\rightarrow \bar{\nu}_{\mu} e^{-} \bar{\nu}_{e} \nu_{\mu}$,
if the four final state leptons equally share the pion energy\citep{Halzen2002}.
The total luminosity of neutrinos emission, which depends on the
distribution of the number of injecting proton $N_{p}(E_{p})$,
can be given as
\be
L_{\nu, 0} = N_{p}(E_{p}) P_{\nu}.
\ee

The primary pions and secondary muons would also undergo
cooling or re-acceleration before they decay into neutrinos.
The final neutrinos luminosity can be estimated
as $L_{\nu} = f_{\rm c} L_{\nu, 0}$, where $f_{\rm c}$ is a
correction factor for cooling or re-acceleration, which
depends on the sources.

The initial neutrino flavor ratio at the source
is $f_{e}:f_{\mu}:f_{\tau}=(\frac{1}{3}:\frac{2}{3}:0)$.
For propagation over cosmological distances,
due to the effect of neutrino oscillations, the
three neutrino flavors approximately mix to the ratio
of (1:1:1) on the earth.

\section{Conclusions and discussion}
Pion production by proton synchrotron was also studied
in semi-classical approximations\cite{Tokuhisa1999,Berezinsky1995}
and in an exact quantum field theory treatment\cite{Maruyama2015, Maruyama2016}.
For $\pi^0$ meson, the result of semiclassical approximations
and our result are basically
on the same order of magnitude near the threshold energy.
Our result has an additional $\pi^0$ hump
via the multi-pion channel in the high energy region,
which is not shown in the semi-classical approximation.

Note that the threshold energy of $\Delta$(1232)-resonance is $\sim$0.2 GeV,
which is less than the rest mass of proton. Therefore, near the
resonance energy,
\be
\gamma_{ p} B/B_{\rm c}^{p} \lesssim 1 ,
\ee
the quantum effect has little influence on the rate of energy loss of
proton \citep{Lieu1993}. Definitely, above the energy equivalent
to the rest mass of proton, the effect of Landau energy levels of
proton should be taken into account.
The quantum result which studied
in more realistic magnetic fields of $10^{15}$ G
and proton energies of $\sim$ TeV, argues that
the anomalous magnetic moment of the proton
can enhance the proton decay width significantly\cite{Maruyama2016}.
In our study, the Fourier analysis of virtual photon spectrum
only considers the time evolution of electric field $\vec{E}(t)$.
As the effect of anomalous magnetic moment is associated with magnetic field,
this effect needs to use different cross section
calculated by the full quantum field theory.
This will be taken into consideration in our future study.

There are several potential astrophysical environments for the
occurrence of the $pB$ process: first,
UHE cosmic rays can be accelerated up to $10^{21}$ eV in
astrophysical sources, including active galactic nuclei, gamma-ray bursts,
starburst galaxies and so on\cite{Hillas1984}.
These cosmic rays can interact with the magnetic field of
white dwarfs ($B \sim 10^{6-9}$ G) in old stellar clusters\cite{Ferrario2015}
in their propagations to the Earth, which
might make the $pB$ process take place.
Second, neutron stars, such as magnetars, could have very strong
magnetic field with strength up-to $B \sim 10^{14-15}$ G.
During the merger of binary neutron stars, high magnetic fields
might be produced
\cite{Duncan1992, Paczynski1992, Usov1992, Price2006},
protons and other heavy elements can be accelerated and interact
with magnetic field in the same system to produce UHE neutrinos
and gamma-ray photons via the $pB$ process.
Third, it is generally believed that it is difficult to produce
UHE photons and neutrinos during the merger of binary black holes,
because there is no conventional targets for the $pp/p\gamma$ interaction.
However, it is proposed that a small amount of gases surrounding
a new-born black hole after the merger of binary black holes could
support a strong magnetic field of $10^{11}$ G in a limited
time\cite{Kotera2016}, which is helpful for the occurrence of
the $pB$ process.

This work is supposed by National Natural Science Foundation of
China (Grant No. 11725312).

%\bibliography{UHEv.bib}

\begin{thebibliography}{10}
\expandafter\ifx\csname url\endcsname\relax
  \def\url#1{\texttt{#1}}\fi
\expandafter\ifx\csname urlprefix\endcsname\relax\def\urlprefix{URL }\fi
\expandafter\ifx\csname href\endcsname\relax
  \def\href#1#2{#2} \def\path#1{#1}\fi

\bibitem{Aartsen2014}
M.~G. {Aartsen}, M.~{Ackermann}, J.~{Adams {\it et al.}}, {Observation of
  High-Energy Astrophysical Neutrinos in Three Years of IceCube Data}, Phys.
  Rev. Letts. 113~(10) (2014) 101101.
\newblock \href {http://arxiv.org/abs/1405.5303} {\path{arXiv:1405.5303}},
  \href {https://doi.org/10.1103/PhysRevLett.113.101101}
  {\path{doi:10.1103/PhysRevLett.113.101101}}.

\bibitem{Fermi1924}
E.~Fermi, {On the Theory of the impact between atoms and electrically charged
  particles}, Z. Phys. 29 (1924) 315--327.
\newblock \href {https://doi.org/10.1007/BF03184853}
  {\path{doi:10.1007/BF03184853}}.

\bibitem{Weizsaecker1934}
C.~v. Weizs{\"a}cker, Ausstrahlung bei st{\"o}{\ss}en sehr schneller
  elektronen, Zeitschrift f{\"u}r Physik 88~(9-10) (1934) 612--625.

\bibitem{Williams1935}
E.~Williams, Ej williams, kgl. danske videnskab. selskab, mat. fys. medd. 13, 4
  (1935), Kgl. Danske Videnskab. Selskab, Mat.-fys. Medd. 13 (1935) 4.

\bibitem{Jackson1998}
J.~D. {Jackson}, {Classical Electrodynamics, 3rd Edition}, John Wiley, 1998.

\bibitem{Lieu1993}
R.~{Lieu}, W.~I. {Axford},
  \href{http://ads.bao.ac.cn/abs/1993ApJ...416..700L}{{Synchrotron Radiation:
  an Inverse Compton Effect}}, Astropys. J. 416 (1993) 700.
\newblock \href {https://doi.org/10.1086/173270} {\path{doi:10.1086/173270}}.
\newline\urlprefix\url{http://ads.bao.ac.cn/abs/1993ApJ...416..700L}

\bibitem{Zhang1998}
J.~L. {Zhang}, Y.~F. {Yuan},
  \href{http://ads.bao.ac.cn/abs/1998ApJ...493..826Z}{The quantum radiation
  formulae of a new radiation mechanism in curved magnetic fields}, Astrophys.
  J. 493 (1998) 826--+.
\newblock \href {https://doi.org/10.1086/305138} {\path{doi:10.1086/305138}}.
\newline\urlprefix\url{http://ads.bao.ac.cn/abs/1998ApJ...493..826Z}

\bibitem{Voisin2017a}
G.~{Voisin}, S.~{Bonazzola}, F.~{Mottez}, {Dirac states of an electron in a
  circular intense magnetic field}, Phys. Rev. D 95~(8) (2017) 085002.
\newblock \href {http://arxiv.org/abs/1703.05193} {\path{arXiv:1703.05193}},
  \href {https://doi.org/10.1103/PhysRevD.95.085002}
  {\path{doi:10.1103/PhysRevD.95.085002}}.

\bibitem{Voisin2017b}
G.~{Voisin}, S.~{Bonazzola}, F.~{Mottez}, {Quantum theory of curvature and
  synchro-curvature radiation in a strong and curved magnetic field, and
  applications to neutron star magnetospheres}, Phys. Rev. D 95~(10) (2017)
  105008.
\newblock \href {http://arxiv.org/abs/1705.03790} {\path{arXiv:1705.03790}},
  \href {https://doi.org/10.1103/PhysRevD.95.105008}
  {\path{doi:10.1103/PhysRevD.95.105008}}.

\bibitem{Tokuhisa1999}
A.~{Tokuhisa}, T.~{Kajino}, {Meson Synchrotron Emission from Central Engines of
  Gamma-Ray Bursts with Strong Magnetic Fields}, \apjl 525 (1999) L117--L120.
\newblock \href {http://arxiv.org/abs/astro-ph/9909286}
  {\path{arXiv:astro-ph/9909286}}, \href {https://doi.org/10.1086/312348}
  {\path{doi:10.1086/312348}}.

\bibitem{Berezinsky1995}
V.~Berezinsky, A.~Dolgov, M.~Kachelrie{\ss}, Curvature radiation by
  ultrarelativistic protons, Physics Letters B 351~(1-3) (1995) 261--265.

\bibitem{Maruyama2015}
T.~Maruyama, M.-K. Cheoun, T.~Kajino, Y.~Kwon, G.~J. Mathews, C.-Y. Ryu,
  Quantum field theoretic treatment of pion production via proton synchrotron
  radiation in strong magnetic fields: Effects of landau levels, Physical
  Review D 91~(12) (2015) 123007.

\bibitem{Maruyama2016}
T.~Maruyama, M.-K. Cheoun, T.~Kajino, G.~J. Mathews,
  \href{http://www.sciencedirect.com/science/article/pii/S0370269316300363}{Pion
  production via proton synchrotron radiation in strong magnetic fields in
  relativistic field theory: Scaling relations and angular distributions},
  Physics Letters B 757 (2016) 125 -- 129.
\newblock \href
  {https://doi.org/https://doi.org/10.1016/j.physletb.2016.03.065}
  {\path{doi:https://doi.org/10.1016/j.physletb.2016.03.065}}.
\newline\urlprefix\url{http://www.sciencedirect.com/science/article/pii/S0370269316300363}

\bibitem{1999PASA...16..160M}
A.~{M{\"u}cke}, J.~P. {Rachen}, R.~{Engel}, R.~J. {Protheroe}, T.~{Stanev},
  {Photohadronic Processes in Astrophysical Environments}, \pasa 16 (1999)
  160--166.
\newblock \href {http://arxiv.org/abs/astro-ph/9808279}
  {\path{arXiv:astro-ph/9808279}}, \href {https://doi.org/10.1071/AS99160}
  {\path{doi:10.1071/AS99160}}.

\bibitem{Muecke2000}
A.~{M{\"u}cke}, R.~{Engel}, J.~P. {Rachen}, R.~J. {Protheroe}, T.~{Stanev},
  {Monte Carlo simulations of photohadronic processes in astrophysics},
  Computer Phys. Commu. 124 (2000) 290--314.
\newblock \href {http://arxiv.org/abs/astro-ph/9903478}
  {\path{arXiv:astro-ph/9903478}}, \href
  {https://doi.org/10.1016/S0010-4655(99)00446-4}
  {\path{doi:10.1016/S0010-4655(99)00446-4}}.

\bibitem{Huemmer2010}
S.~{H{\"u}mmer}, M.~{R{\"u}ger}, F.~{Spanier}, W.~{Winter},
  \href{http://ads.bao.ac.cn/abs/2010ApJ...721..630H}{{Simplified Models for
  Photohadronic Interactions in Cosmic Accelerators}}, Astrophys. J. 721 (2010)
  630--652.
\newblock \href {http://arxiv.org/abs/1002.1310} {\path{arXiv:1002.1310}},
  \href {https://doi.org/10.1088/0004-637X/721/1/630}
  {\path{doi:10.1088/0004-637X/721/1/630}}.
\newline\urlprefix\url{http://ads.bao.ac.cn/abs/2010ApJ...721..630H}

\bibitem{Lieu1995}
R.~{Lieu}, W.~I. {Axford}, {Quantum-limited Synchrotron Radiation in
  Inhomogeneous Magnetic Fields}, \apj 447 (1995) 302.
\newblock \href {https://doi.org/10.1086/175875} {\path{doi:10.1086/175875}}.

\bibitem{Lieu1989}
R.~{Lieu}, J.~J. {Quenby}, W.~I. {Axford}, {Synchrotron radiation treated by
  the Weizsaecker-Williams method of virtual quanta}, Astron. Astrophys. 208
  (1989) 351--356.

\bibitem{Halzen2002}
F.~Halzen, D.~Hooper,
  \href{http://stacks.iop.org/0034-4885/65/i=7/a=201}{High-energy neutrino
  astronomy: the cosmic ray connection}, Rep. on Prog. in Phys. 65~(7) (2002)
  1025.
\newline\urlprefix\url{http://stacks.iop.org/0034-4885/65/i=7/a=201}

\bibitem{Hillas1984}
A.~Hillas, The origin of ultra-high-energy cosmic rays, Annu. Rev. Astron.
  Astrophys. 22~(1) (1984) 425--444.

\bibitem{Ferrario2015}
L.~Ferrario, D.~de~Martino, B.~T. G{\"a}nsicke,
  \href{https://doi.org/10.1007/s11214-015-0152-0}{Magnetic white dwarfs},
  Space Sci. Rev. 191~(1) (2015) 111--169.
\newblock \href {https://doi.org/10.1007/s11214-015-0152-0}
  {\path{doi:10.1007/s11214-015-0152-0}}.
\newline\urlprefix\url{https://doi.org/10.1007/s11214-015-0152-0}

\bibitem{Duncan1992}
R.~C. Duncan, C.~Thompson, Formation of very strongly magnetized neutron
  stars-implications for gamma-ray bursts, Astrophys. J. 392 (1992) L9--L13.

\bibitem{Paczynski1992}
B.~Paczynski, Gb 790305 as a very strongly magnetized neutron star, Acta
  Astron. 42 (1992) 145--153.

\bibitem{Usov1992}
V.~Usov, Millisecond pulsars with extremely strong magnetic fields as a
  cosmological source of gamma-ray bursts, Nat. 357 (1992) 472--474.

\bibitem{Price2006}
D.~J. Price, S.~Rosswog, Producing ultrastrong magnetic fields in neutron star
  mergers, Sci. 312~(5774) (2006) 719--722.

\bibitem{Kotera2016}
K.~{Kotera}, J.~{Silk}, {Ultrahigh-energy Cosmic Rays and Black Hole Mergers},
  Astrophys. J. Letts. 823 (2016) L29.
\newblock \href {http://arxiv.org/abs/1602.06961} {\path{arXiv:1602.06961}},
  \href {https://doi.org/10.3847/2041-8205/823/2/L29}
  {\path{doi:10.3847/2041-8205/823/2/L29}}.

\end{thebibliography}

\end{document}